\documentclass[twocolumn,aps]{revtex4-1}

\usepackage{graphicx} 

\begin{document}

\title{Scalable extraction of error models from the output of error detection circuits}

\author{Austin G. Fowler$^{1,2}$, D. Sank$^1$, J. Kelly$^1$, R. Barends$^1$, John M. Martinis$^1$}

\affiliation{$^1$Department of Physics, University of California, Santa Barbara, California 93106, USA \\
$^2$Centre for Quantum Computation and Communication Technology, School of Physics, The University of Melbourne, Victoria 3010, Australia}

\date{\today}

\begin{abstract}
Accurate methods of assessing the performance of quantum gates are extremely important. Quantum process tomography and randomized benchmarking are the current favored methods. Quantum process tomography gives detailed information, but significant approximations must be made to reduce this information to a form quantum error correction simulations can use. Randomized benchmarking typically outputs just a single number, the fidelity, giving no information on the structure of errors during the gate. Neither method is optimized to assess gate performance within an error detection circuit, where gates will be actually used in a large-scale quantum computer. Specifically, the important issues of error composition and error propagation lie outside the scope of both methods. We present a fast, simple, and scalable method of obtaining exactly the information required to perform effective quantum error correction from the output of continuously running error detection circuits, enabling accurate prediction of large-scale behavior.
\end{abstract}

\maketitle

\section{Introduction}
\label{intro}

The most important open question in the field of quantum computing is whether or not fault-tolerant computation is physically realizable. Answering this question requires, among many other things, measuring the performance of available devices in a manner permitting accurate extrapolation to larger devices. We shall assume a physical system that can be both manipulated with high fidelity and scaled up without changing the physics of the manipulation \cite{Bare13}. In this work, we show how to directly and scalably obtain quantum error correction (QEC) optimized error models for an entire system without performing many separate experiments on small parts of the system. This enables error propagation and composition effects to be fully taken into account, and accurate predictions of large-scale behavior to be made.

Characterizing the performance of quantum gates has a history nearly as long as that of the field of quantum computing itself, which arguably began in 1994 with the publication of Shor's efficient integer factoring algorithm \cite{Shor94b}. A method of completely characterizing a physical process in an open quantum system, namely quantum process tomography (QPT), was first proposed in 1997 \cite{Chua97b,Poya97}. Given $n$ qubits, QPT involves preparing $4^n$ different product states, applying the $n$-qubit process of interest, and measuring in $4^n$ different bases, for a total of $16^n$ different input and measurement combinations, each of which must be repeated many times to obtain a precise average result.

Substantial efforts have been made to improve QPT. Ancilla assisted QPT (AAPT) \cite{Leun00,DAri01,Dur01} replaces the many input states of standard QPT (SQPT) with a single entangled input state, but still involves approximately $16^n$ different measurements. Direct characterization of quantum dynamics (DCQD) \cite{Mohs06} makes use of both entangled input states and Bell measurements to reduce the total work to $4^n$. SQPT, AAPT and DCQD are reviewed and compared in detail in \cite{Mohs08}.

When the process being analyzed corresponds to a sparse matrix in some basis, as is frequently the case in quantum information, compressed sensing techniques can remarkably be used to reduce the number of measurements required to $O(n)$ \cite{Shab11}. The classical postprocessing associated with this procedure remains, however, exponential in $n$. Attempts to reduce the classical postprocessing demands have been made \cite{Flam12}, however the practical and efficient handling of an arbitrarily large number of qubits is not expected, except in special cases \cite{Baum13}. When no specific assumptions can be made on the gate under investigation, at least $2^n$ measurements are required followed by a classical computation of complexity $O(n^2 2^{3n})$ \cite{Reic13}. Improving QPT is an active area of research \cite{Bana13}. Recently, methods incorporating QEC have been proposed \cite{Omka14}.

Despite all of the theoretical advances, fundamentally the output of QPT is a $\chi$-matrix, which is not directly useful in a large-scale QEC simulation. The problem of producing a QEC-appropriate stochastic error model from a $\chi$-matrix is being studied by a number of authors \cite{Gell13,Guti13,Easw13,Puzz14}. However, given an error detection circuit, the approach of performing QPT on each individual gate, ignoring the complexities introduced by all gates working in parallel for an entire error detection cycle, then decomposing each gate $\chi$-matrix into a stochastic error model, is unlikely to enable highly accurate prediction of large-scale behavior.

Another important shortcoming of QPT is the vast number of measurements required to accurately characterize high-fidelity gates. Randomized benchmarking \cite{Emer05,Knil08,Mage11} circumvents this limitation by measuring the average fidelity of a long sequences of gates to accumulate error. When benchmarking a specific gate \cite{Mage12,Kell14}, these sequences consist of alternating random Clifford gates and the gate of interest, terminated by a final gate returning the system to its initial state. The sequence length is chosen such that its total fidelity is less than approximately 0.9, permitting a couple of significant figures of precision to be obtained with a manageable number of different random sequences and repetitions of each sequence. The fidelity of the gate of interest can then be calculated by comparing the obtained alternating sequence fidelity with the fidelity of sequences consisting only of random Cliffords. By tuning gate control parameters for high sequence fidelity, two-qubit gates with fidelities in excess of 0.99 and single-qubit gates with fidelities in excess of 0.999 have been obtained in superconducting qubits \cite{Kell14,Bare13}, sufficiently high to place this technology at the threshold fidelity for surface code quantum computation \cite{Fowl12f}.

Despite the successes of randomized benchmarking, little detail on the structure of gate errors is provided. Furthermore, no information is provided on error propagation and composition within the error detection circuits that a large-scale quantum computer will need to run. An entirely new method, which we describe in this work, is required to scalably and directly obtain the desired QEC-tailored error models.

The discussion is organized as follows. In Section~\ref{rep}, the repetition code is reviewed with an emphasis on the propagation of errors within its error detection circuitry. The repetition code shall be emphasized heavily in this work due to its pedagogical simplicity and immediate experimental accessibility, however the techniques presented are also applicable to the surface code \cite{Brav98,Denn02,Raus07,Raus07d,Fowl12f} and topological cluster states \cite{Fowl09}. These more complex codes are widely believed to offer the greatest hope for realizing a large-scale quantum computer. In Section~\ref{pro}, we describe how to efficiently take the output of repetition code error detection circuits and predict the performance of more complex codes. In Section~\ref{sc}, we discuss the even richer information available if one could directly implement a small surface code. In Section~\ref{super}, we present a discussion of how one could implement our methods using superconducting qubits. Section~\ref{conc} summarizes our results and discusses further work.

\section{The repetition code}
\label{rep}

As its name suggests, the logical states of the repetition code are simply $|0_L\rangle=|00\ldots 0\rangle$ and $|1_L\rangle=|11\ldots 1\rangle$. An arbitrary logical state $|\psi_L\rangle=\alpha|0_L\rangle+\beta|1_L\rangle$ has the property that, in the absence of errors, neighboring qubits have the same value, meaning given $n$ sequentially numbered qubits, measuring an operator $Z_iZ_{i+1}$, $1\leq i<n$, will yield the result +1. A result of -1, or more generally a result that differs from the previous measurement of the same operator, indicates the local presence of an error.

\begin{figure}
\begin{center}
\includegraphics[width=40mm]{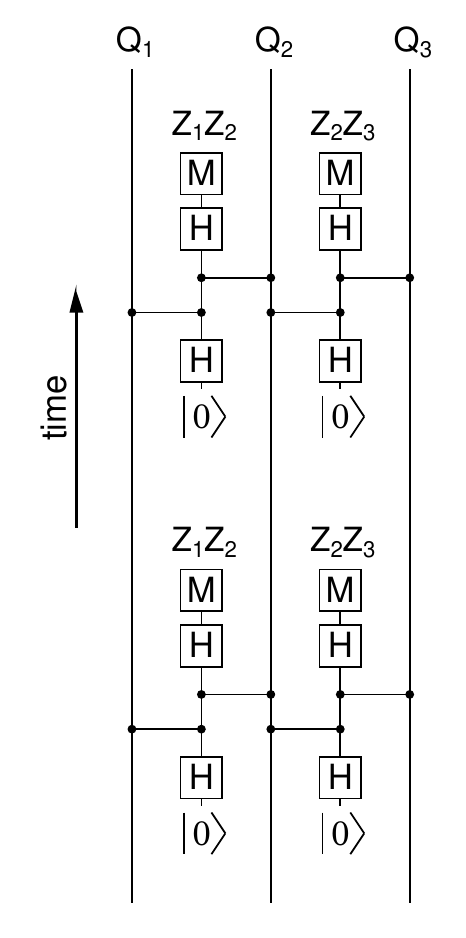}
\end{center}
\caption{Two measurement qubits (repeatedly initialized to $|0\rangle$) interleaved between the 3 data qubits ($Q_1Q_2Q_3$) of a distance 3 repetition code. Two cycles of measuring $Z_1Z_2$ and $Z_2Z_3$ are shown.}\label{rep_circ}
\end{figure}

Measuring $Z_iZ_{i+1}$ operators in a repetition code is straightforward, and can be achieved by inserting a measurement qubit between each data qubit and executing the cyclic circuit of Fig.~\ref{rep_circ}. The measurements in the circuit report 0 for the +1 eigenstate, and 1 for the -1 eigenstate. A \emph{detection event} is defined to be a measurement value that differs from its previous value. Under the assumption that gates only introduce small correlated errors on qubits not directly touched by the gate, there are only a small number of patterns of detection events that can arise from single gate errors. Small magnitude correlated errors on additional qubits are easily handled \cite{Fowl14}, requiring no special processing.

\begin{figure*}
\begin{center}
\includegraphics[width=150mm]{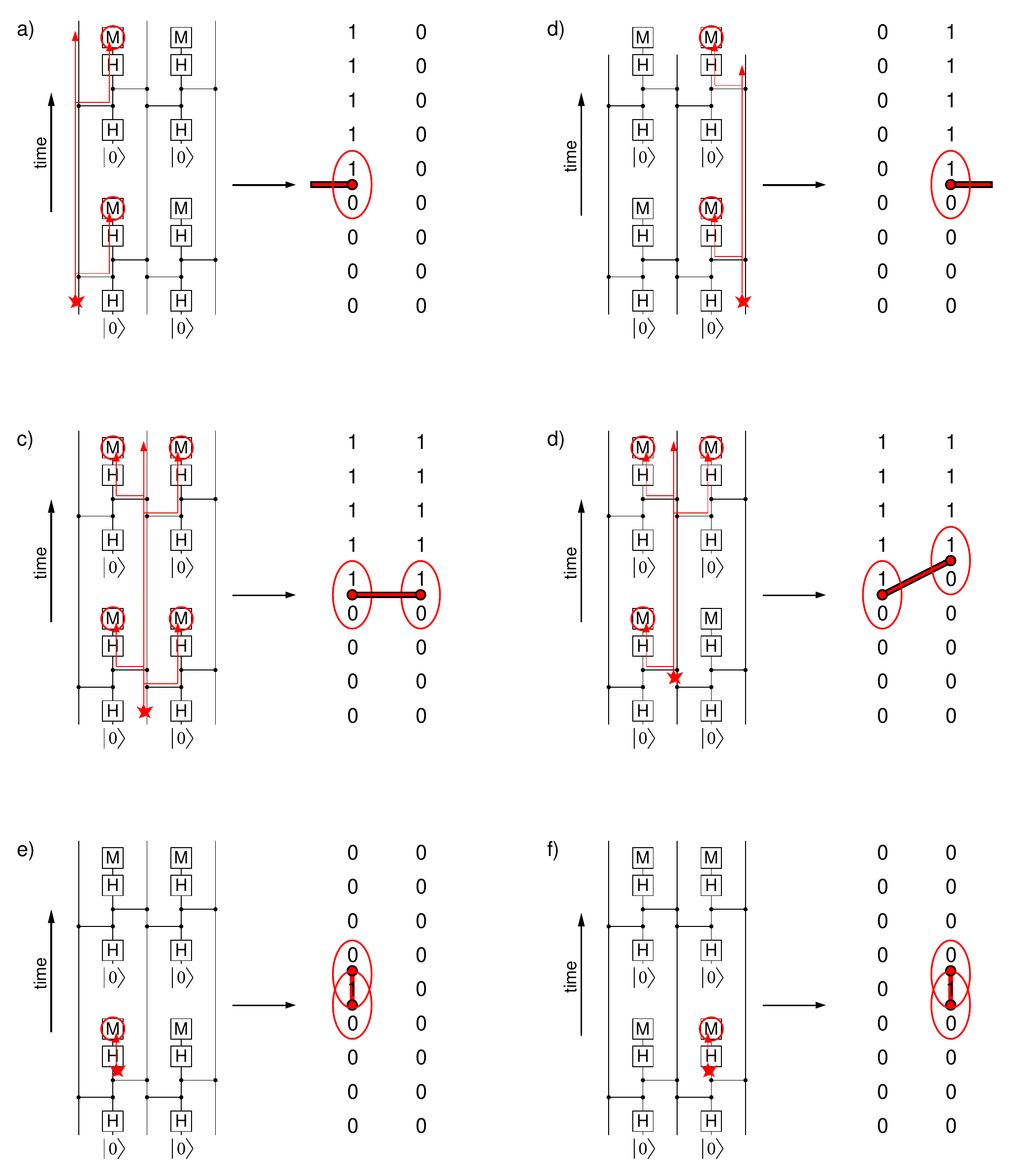}
\end{center}
\caption{a-f) all possible detection event patterns arising from single errors (represented by stars) in a distance 3 repetition code. A detection event (ellipse) is a sequential pair of measurements that differ in value. Errors in the circuit change the measurement values, depicted on the right of each subfigure. The total probability of each possible pattern of detection events is calculated by our open source software Autotune \cite{Fowl12d}. This probability is graphically represented by a cylinder with diameter proportional to this probability (Fig.~\ref{nest}). Cylinder endpoints are placed at the space-time locations of the associated detection events. Cylinders associated with a single detection event a-b) can only occur at the edges of a device.}\label{rep_cases}
\end{figure*}

Figure~\ref{rep_cases} shows every possible detection event pattern arising from single errors in a distance 3 repetition code. Our open source software Autotune \cite{Fowl12d} was used to calculate all possible patterns, and the total probability of each pattern, given, as input, a coded version of Fig.~\ref{rep_circ} and stochastic depolarizing error models for each gate.

Focusing on Fig.~\ref{rep_cases}a, the leftmost data qubit undergoes five identity gates of duration $C_Z$, Hadamard, measurement, initialization and Hadamard  between its own $C_Z$ interactions. An $X$ or $Y$ error during any of these five identity gates will propagate to the neighboring measurement qubit and be detected. Some of the possible errors during the $C_Z$ gate connecting these qubits also contribute to this detection event pattern, specifically $XZ$, $XY$, $YZ$, and $YY$ errors. The total probability of these 14 types of error is the probability of the detection event pattern shown in Fig.~\ref{rep_cases}a.

Working through a second example, in Fig.~\ref{rep_cases}e an $X$ error during initialization, a $Y$ or $Z$ error during the first Hadamard, an $IZ$, $IY$, $ZZ$ or $ZY$ error during the first $C_Z$, a $ZI$, $ZZ$, $YI$ or $YZ$ error during the second $C_Z$, an $X$ or $Y$ error during the second Hadamard, or finally an error during measurement, all contribute to the indicated detection event pattern. The total probability of these 14 types of error is the probability of the detection event pattern shown in Fig.~\ref{rep_cases}e. For reference, every gate and every possible error is exhaustively considered in Appendix~\ref{app}.

\begin{figure}
\begin{center}
\includegraphics[width=50mm]{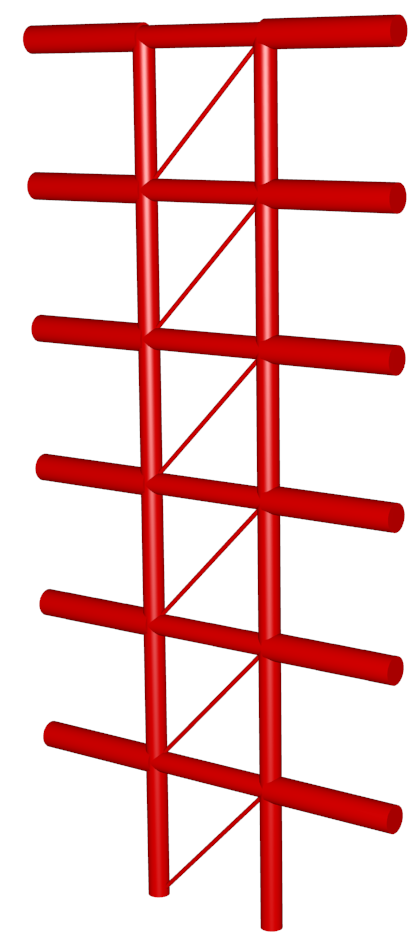}
\end{center}
\caption{Six layers of the nest corresponding to Fig.~\ref{rep_circ}. The diameter of each cylinder is proportional to the probability of detection events at the endpoints of the cylinder. Note that the diagonal cylinder has low probability, due to the small number of errors that can trigger the detection event pattern of Fig.~\ref{rep_cases}d.}\label{nest}
\end{figure}

The information concerning the total probability of different detection event patterns can conveniently be visualized graphically by a 3-D structure of cylinders. The diameter of each cylinder is set to be proportional to the total probability of detection events at the endpoints of the cylinder. We call such a 3-D structure a \emph{nest} \cite{Fowl14}. Six layers of the nest of the repetition code of Fig.~\ref{rep_circ} are shown in Fig.~\ref{nest} --- the cyclic nature of the error detection circuit makes the nest periodic. Note that the vertical cylinders are significantly thicker than the diagonal cylinders, as many more types of error contribute to them.

\section{Processing repetition code error detection output}
\label{pro}

\begin{figure}
\begin{center}
\includegraphics[width=80mm]{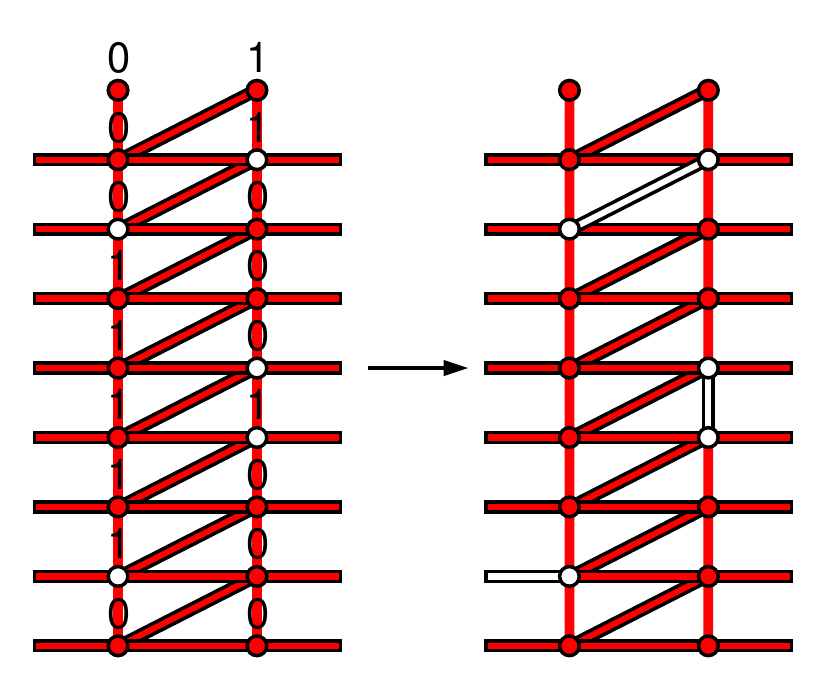}
\end{center}
\caption{(left) Pattern of detection events (white circles). (right) Most probable pattern of errors leading to the observed detection events (white lines).}\label{rep_ex}
\end{figure}

When an error detection circuit is operated well above the threshold fidelity for that circuit, errors, and hence detection events, are sparse. Figure~\ref{rep_ex} gives an example of sparse detection events and the trivial processing required to determine the most probable pattern of errors producing the observed detection events. Clusters of more than two detection events are rare and can be ignored. Clusters of one or two detection events can be matched with each other or to the nearest edge of the device. A running count of the number of times each class of errors is observed can be maintained and, when sufficient statistics have been gathered, the counts can simply be divided by the total number of rounds of error detection to give the probability of each error class.

An arbitrarily large number of qubits can be characterized using an error detection circuit in constant experimental time. The required classical postprocessing is quite trivial and could also be parallelized and performed in constant time. The data generated by our error detection method is precisely that required by our quantum error correction procedure, and by focusing on information obtained only once per round of error detection, all error propagation and composition effects are accounted for.

The repetition code discussed in the previous Section cannot detect all errors. Data qubit $Z$ errors, for example, go unnoticed. See Appendix~\ref{app} for details. A different experiment, such as a four qubit version of the parity measurement experiments described in \cite{Sair14,Chow13}, is required to detect all errors. For example, two measurement qubits and two data qubits arranged in a square, with diagonally opposed measurement qubits measuring operators $ZZ$ and $XX$, are capable of detecting all errors. Nests can be generated for such an experiment using Autotune, in the same manner as has been described for the repetition code.

Given an experimentally determined nest, each cylinder is equivalent to a linear constraint on the sum of specific stochastic terms in the error models of the underlying gates. For example, as discussed in the previous Section, the error class shown in Fig.~\ref{rep_cases}e consists of the total probability of initialization error, a $Y$ or $Z$ error on the first Hadamard, an $X$ or $Y$ error on the second Hadamard, measurement error, and a few other more complex errors that can occur on the $C_Z$ gates. Given the periodic structure of the nest, only a single layer of cylinders needs to be considered. The two nests corresponding to the two different experiments discussed, namely a repetition code and a $2\times 2$  parity check code, therefore give a system of linear equations enabling detailed error models for the underlying gates to be determined.

The central claim of our paper is that error models determined from these experiments will be ideal for accurately estimating the hypothetical performance of a surface code built out of similar gates, providing strong insight into whether arbitrarily reliable quantum computation is technologically feasible.

Even if the actual physical processes leading to the observed detection event patterns do not closely resemble the calculated error models, error correction failure is related only to the density and distribution of detection events. This assertion relies only on low rates of dangerous multi-qubit correlated errors, whose presence or absence can be measured by comparing the experimental logical error rate suppression of the repetition code with that predicted by simulations using the experimentally derived stochastic error models.

\section{Processing surface code error detection output}
\label{sc}

\begin{figure}
\begin{center}
\includegraphics[width=50mm]{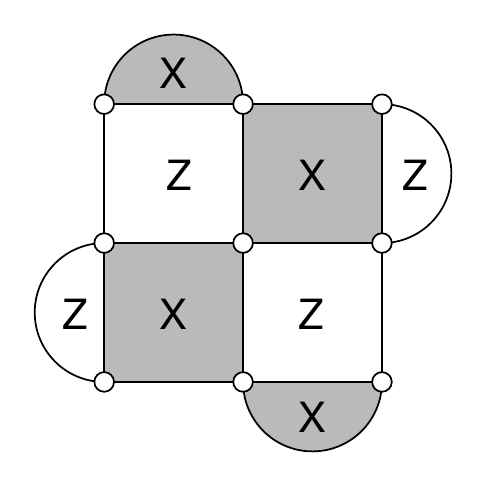}
\end{center}
\caption{The smallest possible surface code, consisting of 9 data qubits. The indicated 8 stabilizers could, in principle, be measured sequentially by a single additional qubit, or measured in parallel by an additional 8 qubits.}\label{sc_fig}
\end{figure}

The smallest possible surface code consists of 9 data qubits plus one or more measurement qubits to handle its 8 stabilizers (Fig.~\ref{sc_fig}). When attempting to predict the performance of a large-scale quantum computer with only nearest neighbor interactions, ideally \cite{Tomi14} a measurement qubit would be devoted to each stabilizer, for a total of 17 qubits.

The surface code contains both $X$ and $Z$ stabilizers, so processing the measurement data of a single cyclic error detection experiment gives two nests, one for each type of stabilizer, that together give complete error information. We will assume that a 17-qubit experiment will not be implemented before gates are significantly above the surface code threshold fidelity. This means we shall continue to assume detection events will be sparse. As we shall see, more information can be obtained from the surface code than was available with the combination of a repetition and $2\times 2$ parity check code.

Consider a single $Y=XZ$ error on the central qubit in Fig.~\ref{sc_fig}. This will lead to both pairs of neighboring $X$ and $Z$ stabilizers changing value. In contrast, a single $X$ or $Z$ error on the central qubit will change only one pair of stabilizer measurements. If $X$ and $Z$ errors are correlated, this will be reflected in detection event correlations between the two nests. Measuring these correlations provides information that can be used advantageously during error correction \cite{Fowl13g}.

Autotune \cite{Fowl12d} can calculate all potential correlations between detection event patterns given a complete set of stochastic error models for each gate. Conversely, given surface code error detection data generated by gates well above threshold, one can calculate the conditional probability of different detection event patterns in the $Z$-stabilizer nest when a given detection event pattern is observed in the $X$-stabilizer nest, and vice versa. This generic experimental approach can be used for any error detection circuit that Autotune handles, which includes topological cluster states, a highly appropriate code for ion traps \cite{Hart14}.

As with the repetition code, the presence of dangerous spatially or temporally correlated errors can be experimentally determined by comparing the observed surface code logical error rate with that predicted by simulations using the error models experimentally determined as described above. Furthermore, an excess probability of experimentally measured spatially or temporarily correlated detection event patterns, higher than that observed in simulations, can be used to specifically identify the unwanted correlated error processes. For example, a defective initialization gate can lead to temporal strings of detection events.

\section{Superconducting qubit example implementation}
\label{super}

The fault-tolerant threshold fidelity of the repetition code is approximately 95\% \cite{Fowl13f}, meaning existing superconducting devices are well above this threshold \cite{Bare13} and therefore well poised to use our presented method. We now discuss some of the implementation details specific to this technology.

Superconducting qubits suffer from leakage to higher-energy states. Generic teleportation-based techniques for handling leakage are known \cite{Fowl13f}, and these techniques reduce leakage to standard gate errors. These leakage-removal techniques, used once per error detection round, effectively enable one to ignore leakage at the circuit level. Without loss of generality, we shall therefore focus on just the implementation of the circuit of Fig.~\ref{rep_circ}.

Running the cyclic circuit of Fig.~\ref{rep_circ} will generate a stream of measurement results from each measure qubit. A single round of error detection is expected to take approximately 500ns. Ideally, we want $10^4$ isolated observations of each detection event pattern shown in Fig.~\ref{rep_cases}. Assuming a universal set of gates with error rates slightly below 1\%, the total probability of each detection event pattern should be a few percent, meaning less than $10^6$ rounds of error detection should be required. This is expected to be possible in under one second.

By searching for isolated detection event patterns in the data, the repetition code nest can be constructed and the performance of the code simulating using the derived stochastic error models. By comparing this simulated performance with the observed logical error rate performance of the repetition code, rich information on the presence or absence of dangerous correlated errors can be inferred.

\section{Discussion}
\label{conc}

We have described a method of producing exactly the information required for efficient quantum error correction directly from the output of error detection circuits. Our method could in principle be used while a large-scale quantum computer was running, to dynamically monitor the performance of the hardware. Repetition code experiments have been described with the potential to accurately predict the performance of a surface code quantum computer, thereby potentially demonstrating that arbitrarily reliable quantum computation is technologically feasible.

Superconducting qubit experimental work making use of our proposed method is in preparation. We expect to be able to obtain error information accurate to 1\% in under one second, with the time independent of the number of qubits characterized. Classical postprocessing requirements are trivial, and could also be parallelized to constant cost.

As more quantum technologies surpass threshold fidelities for various codes, we believe characterization methods based on error detection circuits will become standard. Rapid individual gate techniques based on randomized benchmarking still remain important for tuning new devices \cite{Kell14}.

\section{Acknowledgements}
\label{ack}

We thank Joydip Ghosh, Daniel Puzzuoli, and Evan Jeffrey for helpful discussions. This research was funded by the US Office of the Director of National Intelligence (ODNI), Intelligence Advanced Research Projects Activity (IARPA), through the US Army Research Office grant No. W911NF-10-1-0334. Supported in part by the Australian Research Council Centre of Excellence for Quantum Computation and Communication Technology (CE110001027) and the U.S. Army Research Office (W911NF-13-1-0024). All statements of fact, opinion or conclusions contained herein are those of the authors and should not be construed as representing the official views or policies of IARPA, the ODNI, or the US Government.

\bibliography{../References}

\appendix

\section{Exhaustive tabulation of errors}
\label{app}

\begin{figure}
\begin{center}
\includegraphics[width=70mm]{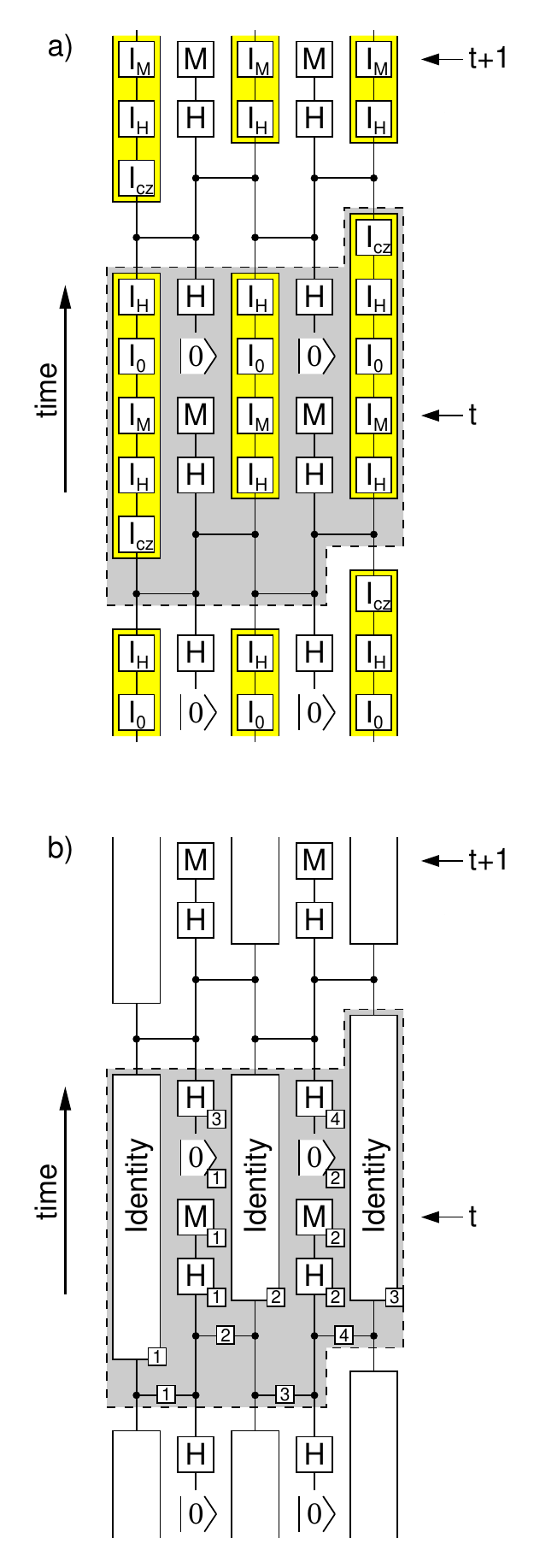}
\end{center}
\caption{a) Detailed repetition code quantum circuit designed to measure $ZZ$ operators. b) Simplified and labeled quantum circuit with three long identity gates.}\label{rep_detailed}
\end{figure}

Here we tabulate every error on every gate in the repetition code, and indicate where and when these errors are detected. Note that not all errors are detected. The repetition code involves measuring $ZZ$ operators, and the quantum circuit of the three data qubit version of this code is shown in Fig.~\ref{rep_detailed}a. For brevity, the sequences of identity gates have been collapsed to single identity gates in Fig.~\ref{rep_detailed}b, which also has each gate numbered.

We focus on pure $X$ and $Z$ errors, shown in Table~\ref{Tab}. A pattern of detection events corresponding to combinations of these errors can be trivially computed. For example, an $XZ$ error on gate $C_{Z1}$ means both an $XI$ error and an $IZ$ error leading to cancellation of the $L_{t+1}$ terms and a single $L_{t}$ detection event.

\begin{table}
\begin{tabular}{c|c|c|c|c}
Error & DE1 & DE2 \\
\hline
$C_{Z1}(IX)$ & - & - \\
$C_{Z1}(XI)$ & $L_{t+1}$ & - \\
$C_{Z1}(IZ)$ & $L_{t}$ & $L_{t+1}$ \\
$C_{Z1}(ZI)$ & - & - \\
\vspace{-2mm} \\
$C_{Z2}(IX)$ & $L_{t+1}$ & $R_{t+1}$ \\
$C_{Z2}(XI)$ & - & - \\
$C_{Z2}(IZ)$ & - & - \\
$C_{Z2}(ZI)$ & $L_{t}$ & $L_{t+1}$ \\
\vspace{-2mm} \\
$C_{Z3}(IX)$ & - & - \\
$C_{Z3}(XI)$ & $L_{t}$ & $R_{t+1}$ \\
$C_{Z3}(IZ)$ & $R_{t}$ & $R_{t+1}$ \\
$C_{Z3}(ZI)$ & - & - \\
\vspace{-2mm} \\
$C_{Z4}(IX)$ & $R_{t+1}$ & - \\
$C_{Z4}(XI)$ & - & - \\
$C_{Z4}(IZ)$ & - & - \\
$C_{Z4}(ZI)$ & $R_{t}$ & $R_{t+1}$ \\
\vspace{-2mm} \\
$I_1(X)$ & $L_{t+1}$ & - \\
$I_1(Z)$ & - & - \\
\vspace{-2mm} \\
$I_2(X)$ & $L_{t+1}$ & $R_{t+1}$ \\
$I_2(Z)$ & - & - \\
\vspace{-2mm} \\
$I_3(X)$ & $R_{t+1}$ & - \\
$I_3(Z)$ & - & - \\
\vspace{-2mm} \\
$H_1(X)$ & $L_{t}$ & $L_{t+1}$ \\
$H_1(Z)$ & - & - \\
\vspace{-2mm} \\
$H_2(X)$ & $R_{t}$ & $R_{t+1}$ \\
$H_2(Z)$ & - & - \\
\vspace{-2mm} \\
$H_3(X)$ & - & - \\
$H_3(Z)$ & $L_{t+1}$ & $L_{t+2}$ \\
\vspace{-2mm} \\
$H_4(X)$ & - & - \\
$H_4(Z)$ & $R_{t+1}$ & $R_{t+2}$ \\
\vspace{-2mm} \\
$M_1(X)$ & $L_{t}$ & $L_{t+1}$ \\
$M_2(X)$ & $R_{t}$ & $R_{t+1}$ \\
$|0\rangle_1(X)$ & $L_{t+1}$ & $L_{t+2}$ \\
$|0\rangle_2(X)$ & $R_{t+1}$ & $R_{t+2}$ \\
\end{tabular}
\caption{Where and when each possible error is detected, if it is detected. For example, a $Z$ error on Hadamard gate $H_3$ is detected by the left measurement qubit during measurement rounds $t$+1 and $t$+2. An $X$ error on Identity gate $I_2$ is detected by the left and right measurement qubits during round $t$+1. Two-qubit errors are ordered left to right, so an $XI$ error on $C_{Z3}$ means an $X$ error on the central data qubit, which is detected by the left measurement qubit in round $t$ and the right measurement qubit in round $t$+1.}
\label{Tab}
\end{table}

\end{document}